\documentclass[a4,twocolumn]{aastex63}
\usepackage{amsmath}
\usepackage{natbib}
\usepackage[T1]{fontenc}

\shorttitle{GRB Host Galaxies}
\shortauthors{Bj\"ornsson}

\begin{document}

\title{Gamma-Ray Burst Host Galaxies:\\Specific Star Formation Rate vs.~Metallicity}

\author[0000-0003-1800-6382]{Gunnlaugur Bj\"ornsson}
\affiliation{Centre for Astrophysics and Cosmology\\Science Institute, University of Iceland, Dunhagi 5, 107 Reykjavik, Iceland}
\email{gulli@hi.is}

\begin{abstract}
    The observed properties of long gamma-ray burst (GRB) host galaxies show them to often be of a rather low metallicity and/or of high specific star formation rate (SFR). It is not clear which of these properties is a dominant factor in determining if a galaxy will host a GRB or not. In fact there are indications, at least in the local Universe, that the two may be anticorrelated and that the metallicity is the deciding parameter. Here, we consider GRB production models dependent on both
    quantities and show that when compared to the best available data, the respective star formation fractions appear indistinguishable out to redshift of $z\sim 4$. However, the fraction of galaxies hosting a GRB, as determined by the specific SFR, is less at tension with the observed host galaxy fraction than the corresponding metallicity determined fraction, but this conclusion is model dependent. Well established galaxy stellar mass and star formation rate functions at high redshift are crucial in breaking the apparent degeneracy between the specific rate and metallicity in GRB production probability.  
\end{abstract}

\keywords{galaxies: evolution -- galaxies: high-redshift -- galaxis: star formation --
gamma-ray bursts: general}

\section{Introduction}

Long duration gamma-ray bursts (GRBs) are an important and unique tool
in cosmological studies. They have been shown to be associated with
the final moments in the lives of massive stars  \citep{Woosley1993, WB2006}, and the
formation of core-collapse supernovae \citep[e.g.,][]{Hjorth2003, Stanek2003}. 
The rate of their formation has been suggested to trace the cosmic star formation 
rate (SFR) and thus probe the star formation history of the early universe \citep[e.g.][]{Chary2016}.

It has, however, become clear that GRBs are not as un\-biased as tracers of star 
formation as was previously thought, and the comoving SFR density traced by GRBs is modestly higher than the comoving SFR density at high redshifts 
\citep[e.g.][]{Kistler2008, RE2012, Perley2016b}. A thorough understanding of 
the GRB host galaxy contribution to the cosmic SFR is fraught 
with selection effects and evolutionary uncertainties at high redshifts.
 
The subpopulation of GRB-selected galaxies may be prone to hitherto unknown or unidentified biases, making their relation to typical star-forming galaxies difficult to quantify. A carefully selected and a complete sample of GRB host galaxies is therefore essential to guarantee as far as possible an unbiased view of the host population. \cite{Vergani2015} and \cite{Japelj2016} used a complete sample of 58 host galaxies from the \textit{Swift}/BAT6 to study the low-redshift host population ($z<1$), while \cite{Palmerio2019} extended that to $1<z<2$. Recently, results of two high-redshift host galaxy surveys have also been published: TOUGH \citep{Hjorth2012, Jakobsson2012, Kruhler2012, Michalowski2012, Milvang-Jensen2012,Schulze2015} and SHOALS \citep{Perley2016a, Perley2016b}. SHOALS is a larger sample (119 hosts), and selected using different criteria from TOUGH (69 hosts), but their redshift  distributions are essentially identical as discussed by \cite{Perley2016b}. In fact, SHOALS is the largest, most redshift-complete, unbiased host galaxy sample available and extends out to $z\gtrsim 6$.

It has been suggested that metallicity dominates over specific SFR (SFR per unit stellar mass, hereafter denoted by $S^*$), as the deciding factor for the probability of a galaxy hosting a GRB or not. Host galaxies of GRBs tend to have a high $S^*$ \citep[e.g.][]{Christensen2004, Savaglio2009, Perley2013, Hunt2014}, while many are found also to be of low metallicity \cite[e.g.][]{Modjaz2019, Palmerio2019}. In fact, \cite{Modjaz2019} find that the hosts of their (low-redshift) sample are of low
metallicity and high $S^*$, while \cite{Palmerio2019} reach a similar conclusion for $1<z<2$. In addition, \cite{Modjaz2019} argue that since the $S^*$ vs.~metallicity relation of supernova-GRB hosts lies well below the corresponding relation for SDSS galaxies, the metallicity is the more likely fundamental factor in GRB production, at least for $z<0.16$. Both papers conclude that metallicity determines the GRB production efficiency within the redshift range they consider. 
 
\cite{Perley2016a, Perley2016b}, however, find an essentially redshift-independent metallicity threshold in their high-redshift SHOALS sample ($z\lesssim 6.3$), close to solar ($\sim 0.5 Z_\sun$ to $\sim Z_\sun$). In addition, \cite{Perley2016b} apply a luminosity threshold to their host galaxy sample and show that the fraction of star-formation contributed by the host galaxies below this threshold increases with increasing redshift.

In this paper, we estimate the fractional contribution of GRB host galaxies to the global star formation as a function of redshift. For this purpose, we adopt commonly used galaxy SFR-stellar-mass relations, galaxy stellar-mass functions and a mass-metallicity relation. Although many GRB host galaxies have been shown to have a high $S^*$, as discussed above, the potential importance of this parameter in the context of GRB formation has not been explored. Our main emphasis is therefore on the
role of a specific SFR threshold therein, but we also consider the relative importance of the $S^*$ and metallicity thresholds in determining the likelihood of a galaxy hosting a GRB.
Starburst galaxies are not considered in this work. Their contribution to the cosmic SFR density has been shown to be small, on the order of $10-15\%$ \citep{Rodighiero2011, Sargent2012}, and not to evolve with time \citep{Schreiber2015}, see also \cite{Caputi2017} for higher redshift results.

Section \ref{ingredients} presents our models and assumptions while Section~\ref{rest}  
outlines our main results. The paper concludes with Section~\ref{conclusions}. Throughout, we assume a standard cosmology with $\Omega_m=0.3$,
$\Omega_\Lambda=0.7$ and $H_0=70$ km s$^{-1}$Mpc$^{-1}$.

\section{The GRB-SFR Relation}
\label{ingredients}

We follow the approach of \cite{RE2012} and consider scenarios where the rate of GRB formation is proportional to the SFR density, the proportionality expressed as 
a redshift-dependent function, $\Psi(z)=\Psi_0\psi(z)$. The constant $\Psi_0$ represents the number of GRBs produced per unit stellar mass and $\psi(z)$ is the fraction of star formation that can produce a GRB at redshift $z$. We further assume that the probability of a GRB being formed in a galaxy is most strongly determined by a 
single basic parameter, e.g.,~metallicity or specific SFR, $S^*$. We implement
this by assuming that for a metallicity above a given threshold value, GRB formation
will be suppressed, while for $S^*$, we assume that to host a GRB, a galaxy will need to have a specific rate above a specified threshold value. The metallicity threshold has previously
been considered by, e.g.~\cite{Virgili2011} and \cite{RE2012}, but a specific SFR threshold is, for the first time, applied here to models exploring GRB host galaxy contribution to cosmic star formation.

\subsection{Metallicity}

As in \cite{RE2012}, we first assume that GRB production will be suppressed if the metallicity is above a specified ceiling, $12+\log[O/H]_{\rm crit}$ on the \cite{KK2004} scale. For ease of comparison with the results of \cite{RE2012}, we also adopt the redshift dependent stellar-mass-metallicity relation as parameterized by \cite{Savaglio2005},

\begin{align}
12+\log[O/H]_{\rm crit}&=-7.5903+2.5315\log M_{\rm cr}\nonumber\\
&-0.09649\log^2 M_{\rm cr}\nonumber\\
&+5.1733\log t_H -0.3944\log^2 t_H \nonumber\\
&-0.403\log t_H \log M_{\rm cr}.
\label{eq:met}
\end{align}
Here, $t_H$, is the Hubble time and we denote the critical galaxy stellar mass corresponding to a metallicity ceiling by $M_{\rm cr}$, above which GRB production is suppressed.  Obtaining an accurate mass-metallicity relation at high redshift is observationally challenging and requires different galaxy selection methods and different metallicity estimators than at lower redshift \citep{Maiolino2019}. For the purpose of this work, we assume the above expression is applicable over the redshift range we consider.

\subsection{Galaxy Stellar Mass Function}

We next assume that the galaxy stellar-mass function can be analytically expressed with a
Schechter-type function \citep{DA2008}:

\begin{equation}
\Phi(M,z)dM=\Phi_0\left(M/M_1\right)^{\gamma}\exp(-M/M_1)\frac{dM}{M_1},
\label{eq:phi}
\end{equation}
where the normalization, $\Phi_0$, the characteristic mass, $M_1$, and the
low mass slope, $\gamma$, in general all depend
on the redshift. We adopt the parameters derived
in \cite{DA2008},
$\Phi_0(z)\approx 0.003(1+z)^{-1.07}{\rm Mpc}^{-3}{\rm dex}^{-1}$,
$\log[M_1/M_{\sun}](z)\approx 11.35-0.22\ln(1+z)$, and we take
$\gamma \approx -1.3$, independent of $z$. \cite{Fontana2006} derived
a different $z$-dependence of all three parameters, with $M_1$ and
$\gamma$ important for our discussion. Recent work has shown that the 
stellar-mass function may be better represented by a double Schechter 
function \citep[active and passive galaxies, e.g.][]{Tomczak2014, Davidzon2017}, 
although \cite{Tomczak2014} find that a single Schechter function is sufficient for $z>2$. 
Our approach amounts to assuming that only the active population contributes to GRB production.

\subsection{Specific SFR}

Finally, as the remaining model ingredient, we adopt the analytic SFR-stellar-mass relation of \cite{DA2008},

\begin{equation}
   SFR(M,z)=S_0\left(M/M_0\right)^{\beta}\exp(-M/M_0),
   \label{eq:sfrm}
\end{equation}
where the low mass slope is fixed at $\beta\approx 0.5$, the 
redshift-dependent SFR is $S_0(z)=3.01(1+z)^{3.03}M_{\sun}{\rm
  yr}^{-1}$ and the characteristic mass is given by $M_0(z)=2.7\times
10^{10}(1+z)^{2.1}M_{\sun}$. 

Using equation~(\ref{eq:sfrm}), we define the galaxy specific SFR-stellar-mass relation simply as
\begin{align}
S^*\equiv SFR/M =(S_0/M_0)(M/M_0)^{\beta-1}\exp(-M/M_0).
\label{eq:ssfr}
\end{align}
We express the specific rate in Gyr$^{-1}$, and again denote by $M_{\rm cr}$ the critical mass corresponding to a given specific rate threshold, \textit{below} which GRB production is suppressed, i.e.\ we require a galaxy to have a specific rate {\em above} a threshold value to produce a GRB. The redshift dependence of $S^*$ for a fixed mass $M<M_0$, is most strongly determined by the ratio, $S_0/M_0 \sim (1+z)$, and $(1/M_0)^{\beta -1}\sim M_0^{1/2}\sim(1+z)$, hence, $S^*\sim(1+z)^2$, in good agreement with 
high-redshift data \citep{Duncan2014,Salmon2015,Bhatawdekar2018}.

\begin{figure}
	\includegraphics[scale=0.46]{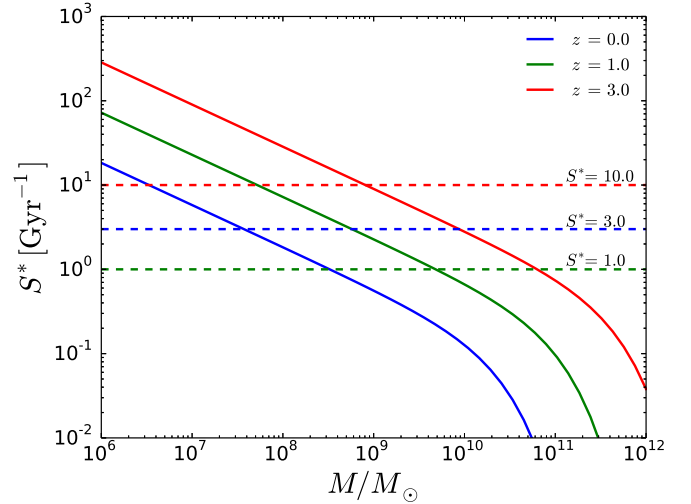}
	\caption{Specific star formation rate, $S^*$, from equation~(\ref{eq:ssfr}), as a function of stellar mass for three different redshifts, $z=0,1,$ and $3$, (from the bottom up). Also plotted as horizontal dashed lines are the constant specific rate values of $S^*=1,3,$ and $10$ Gyr$^{-1}$ (also from the bottom up). Note that a given specific rate corresponds to a higher galaxy mass as the redshift increases.}
	\label{fig:ssfr_mass}
\end{figure}

In Figure~\ref{fig:ssfr_mass}, we plot $S^*$ as a function of mass at three different redshifts. We also indicate three different specific rate values that we explore further below. Note that $S^*$ is a monotonically decreasing function of mass, as $\beta-1\approx -0.5$. This implies that if galaxies with high $S^*$ are on average more likely to host a GRB than galaxies with lower specific rates, they are also on average of lower mass and thus of lower luminosity for most observed galaxy mass to
light ratios. The condition of a high $S^*$ on GRB hosts, therefore, also implies that, on average, they are of low luminosity. Furthermore, a specific rate threshold,  \textit{below} which GRB production is suppressed translates into a critical mass threshold \textit{above} which GRB production is suppressed.

\subsection{Critical Mass Limits}
\label{mcr}
Our assumption that the production probability of a GRB in a given host galaxy hinges on a single parameter characterizing an average property of the galaxy allows us to isolate its effects and explore its consequences. Here, we will discuss two criteria.

The first is the metallicity threshold, already mentioned, that provides an upper mass limit on galaxies that will be able to host a GRB (see equation~\ref{eq:met}).

The second criterion we explore is that a galaxy will be more likely to host a GRB if it has a high specific SFR. We implement this by assuming that to host a GRB, a galaxy will need to have a specific rate above a given threshold value that we further assume to be independent of redshift. Again, this provides an upper stellar-mass limit for galaxies hosting GRBs. 

\begin{figure}
	\includegraphics[scale=0.46]{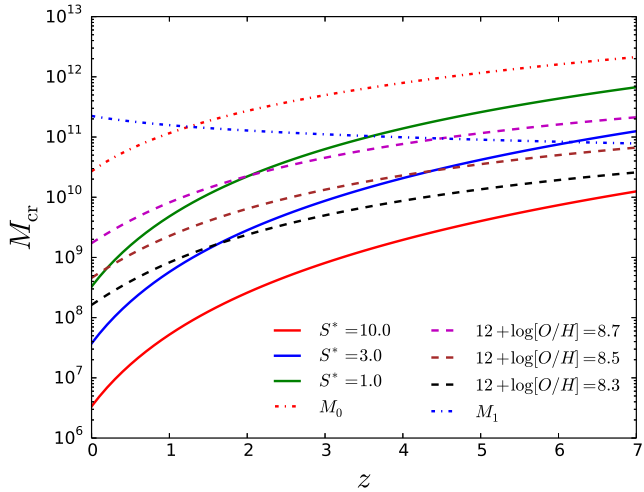}
	\caption{Critical galaxy stellar mass (in solar masses) as a function of redshift, $z$, for three different values of $S^*$ (in Gyr$^{-1}$, solid curves) and three different metallicity thresholds (dashed curves). The critical mass increases faster with increasing redshift for the $S^*$ thresholds. For galaxies more massive than the critical mass, GRB production is suppressed. For comparison we also show the redshift evolution of $M_0$ and $M_1$ (dashed-dotted).}
	\label{fig:mcr}
\end{figure}

Figure~\ref{fig:mcr} shows examples of the critical masses as a function of redshift for three different values of $S^*$ thresholds, as well as three different metallicity thresholds. The critical mass increases more rapidly with increasing redshift for the $S^*$ thresholds, than for the metallicity thresholds. In addition, at a fixed redshift, the critical mass \textit{increases} with an increasing metallicity threshold, but \textit{decreases} with an increasing specific rate threshold. Galaxies
with masses above the critical mass values have, in this model, a very low probability of producing a GRB.

It is clear that for both threshold types ($S^*$ and metallicity), in fact, very few if any GRBs would be produced by galaxies in the mass range $10^{10}-10^{11} M_\sun$ out to redshifts $z \lesssim 2$. Indeed, galaxies of mass $\sim 10^{11} M_\sun$ would not be hosting a GRB out to a redshift of $z\sim 3-4$, in either scenario, as observed by \cite{Perley2016b}.

Only in the case of relatively low values of $S^*$ and rather high values of the metallicity threshold does the critical mass reach the characteristic mass, $M_1$, of equation~(\ref{eq:phi}), at the highest redshifts shown. Massive galaxies, around  $M_1$ or higher, are only likely to host a GRB at high redshift and thus contribute to the GRB-probed star formation if of high metallicity or of low specific SFR. 

\section{The SFR Fraction}
\label{rest}

We now proceed as in \cite{RE2012} and define the fraction of star formation occurring in galaxies with $M<M_{\rm cr}$ as:

\begin{equation}
\psi(z)=\frac{\int_0^{M_{\rm cr}} SFR(M,z)\Phi(M,z)dM}{\int_0^\infty SFR(M,z)\Phi(M,z)dM}=P(1+\beta+\gamma,t_{\rm cr}).
\label{eq:psi}
\end{equation}
Here, $P(x,y)=\Gamma(x,y)/\Gamma(x)$, with $\Gamma(x)$ the gamma function,

\begin{equation}
\Gamma(x,y)=\int_0^y t^{x-1}e^{-t}dt,
\end{equation}
is the lower incomplete gamma function, and the second equality in eq.~(\ref{eq:psi}) holds for the functional forms adopted for $\Phi(M,z)$ and $SFR(M,z)$. Recall that $\psi(z)$ is also the fraction of star formation producing a GRB \citep{RE2012}.

In equation~(\ref{eq:psi}), $t_{\rm cr}=M_{\rm cr}(1/M_0+1/M_1)$, and
the redshift dependence of $\psi(z)$ is only through the mass scales if the power-law
exponents $\beta$ and $\gamma$ are redshift independent. 

\begin{figure*}
	\includegraphics[scale=0.5]{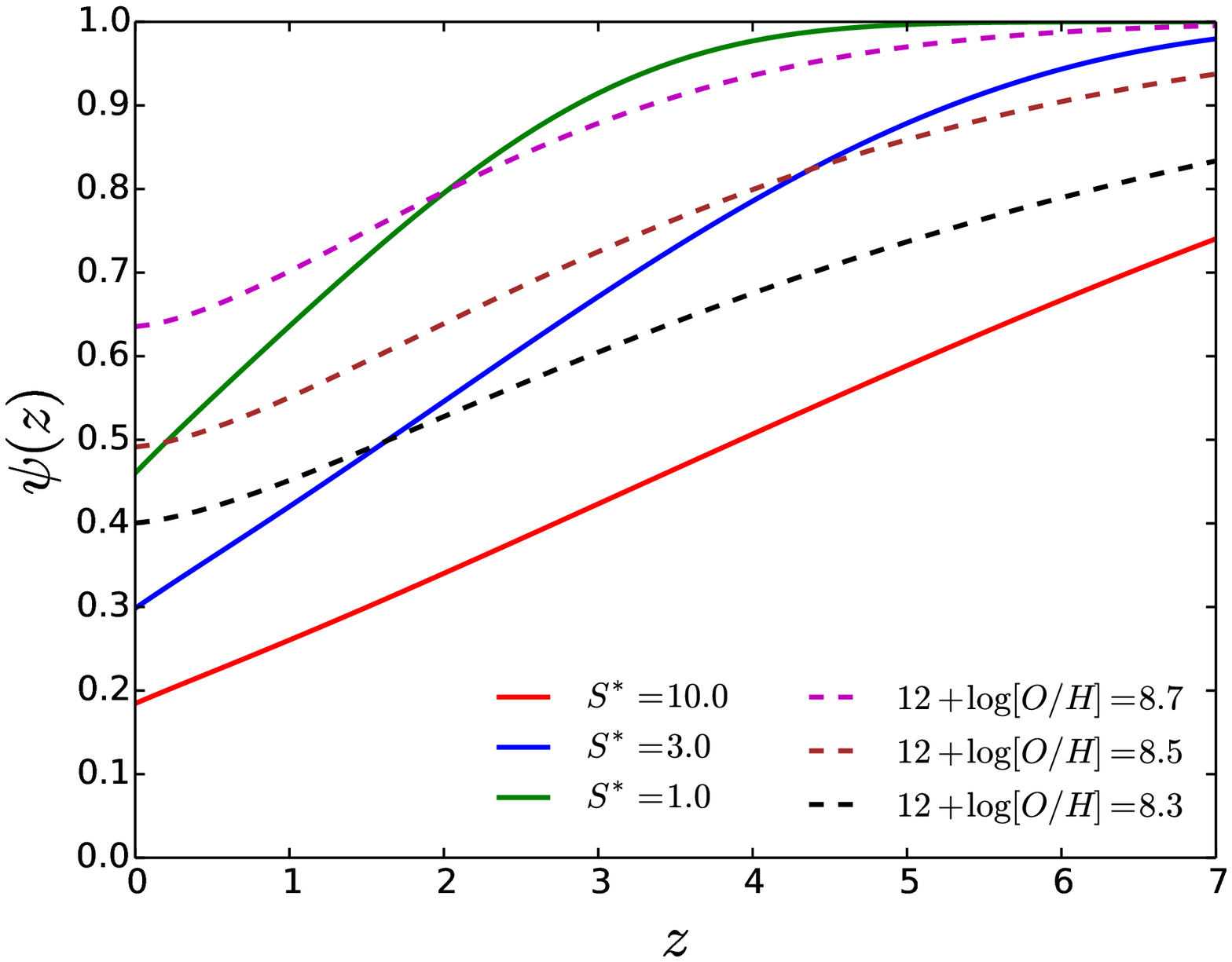}	
	\includegraphics[scale=0.5]{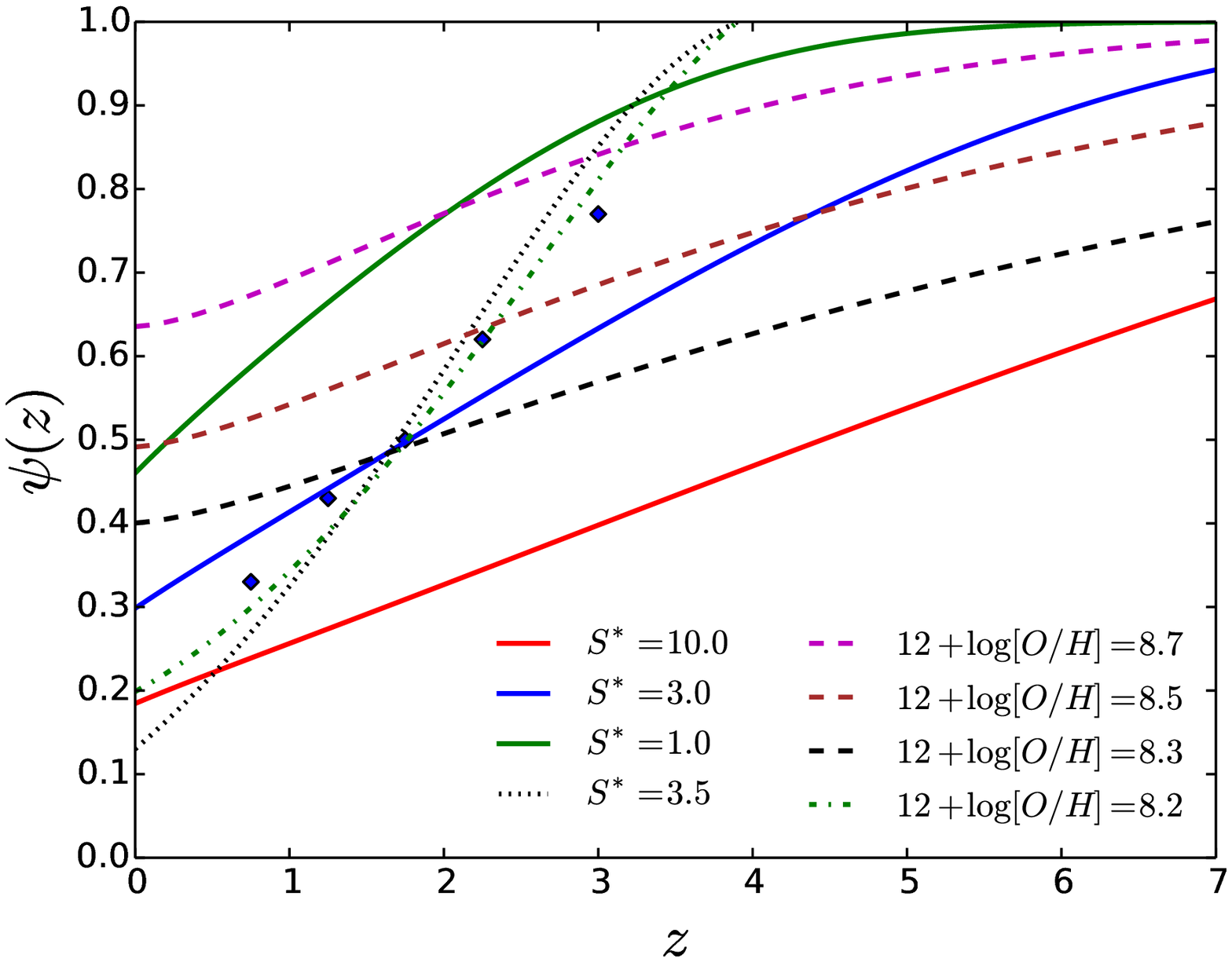}	
	\caption{(Left panel) Fraction of star formation in galaxies with stellar mass $M < M_{\rm cr}$ (equation~\ref{eq:psi}), as a function of redshift, $z$. Solid curves are for specific star formation rate thresholds, while the dashed curves are for different metallicity thresholds. (Right panel) Diamonds show the fraction of star formation contributed by GRB host galaxies, as determined by \cite{Perley2016b}, plotted at the center of their redshift bins. The black dotted ($S^*$) and green
    dashed-dotted (metallicity) curves show the fractions obtained using the fit of \cite{Fontana2006} to $\gamma$ and $M_1$. See text for discussion.
	}
	\label{fig:psi}
\end{figure*}

In Figure~\ref{fig:psi} (left panel), we plot the evolution of this fraction as a function of redshift for the same threshold values as in Fig.~\ref{fig:mcr}, both for $S^*$ (solid curves), and metallicity (dashed curves).  In general, the {\em lower} the metallicity threshold, the lower the star formation fraction in low mass galaxies at a given redshift. In contrast, the {\em higher} the specific rate threshold,  the lower the star formation fraction in low mass galaxies at a fixed redshift. This is a direct consequence of the  redshift dependence of $M_{\rm cr}$ for the two different threshold criteria, as discussed above. 

It is clear that the specific rate thresholds translate into a stronger redshift dependence of the corresponding star formation fractions, $\psi(z)$, than is the case for the metallicity thresholds. This, again, is inherited through the redshift dependence of $M_{\rm cr}$. Hence, the higher the specific SFR, the lower the contribution of these galaxies to the total star formation at all redshifts. As an example, at redshift $z=1$, the star formation fraction would be over $0.6$ for $S^*=1$,
while only about $0.25$ for $S^*=10$. In general, therefore, an increasing specific SFR threshold suppresses the GRB formation rate at a given redshift more strongly than a decreasing metallicity threshold does. As a consequence, the $S^*$ and metallicity thresholds may appear anticorrelated if viewed this way, but we emphasize that in the models discussed here, there is no direct physical connection between $S^*$ and the metallicity thresholds.

We plot the data points from \citet[Table 2]{Perley2016b} on top of Fig.~\ref{fig:psi} (right panel). These show the fraction of cosmic star formation contributed by GRB hosting galaxies at each epoch, as inferred from the SHOALS sample. These fractions are determined by imposing a luminosity threshold above which GRB formation is suppressed, which directly corresponds to our $M_{\rm cr}$ thresholds. The SHOALS fractions rise more steeply with $z$, than both the metallicity and the $S^*$ fractions presented here. However, the redshift evolution of the latter is more in line with the SHOALS data, as a comparison with the $S^*=3.0$ curve shows, although it does not rise sufficiently steeply. 

Replacing our constant $\gamma$ and parameterization of $M_1$ in the galaxy stellar-mass function (eq.~\ref{eq:phi}) with the corresponding fits from \cite{Fontana2006}, leaving other model ingredients unchanged, we plot the resulting fractions with the black dotted and green dashed-dotted curves in Fig.~\ref{fig:psi}. We emphasize that this is not a fit, but an overplot of the star formation fraction using a different parameterization for the galaxy stellar-mass function and varying the
metallicity and $S^*$ threshold values until a reasonable by-eye agreement is obtained. Overall it gives a better resemblance to the data for both types of thresholds, but we remark that the \cite{Fontana2006} fit is restricted to $z\lesssim 4$. This implies that the star formation fraction is more strongly dependent on the galaxy relations than the nature of the imposed thresholds. As indicated in the figure, an $S^*$ threshold value of $3.5$  Gyr$^{-1}$ is sufficient to account for the
behavior. Specific rates of the SHOALS galaxies are not available, and our threshold value is somewhat higher than the median value of $0.8$ Gyr$^{-1}$ in the host sample of \cite{Savaglio2009}, and the typical value of $S^*\approx 1$ Gyr$^{-1}$ found by \cite{Perley2013} and \cite{Hunt2014}. Note also that the data shows a shallower redshift evolution than both the metallicity and $S^*$ threshold curves, with the former tracing that behavior better than the latter. However, a low value of the
metallicity threshold ($12+\log[O/H]=8.2$), is needed to accommodate the data for this parameterization of the mass function, but apparently too low to be consistent with the inferred metallicity threshold of the SHOALS data \citep[and other recent estimates, e.g.][]{Palmerio2019}. This indicates that the redshift behavior of the host galaxy star formation fraction is sensitive to the galaxy stellar-mass function parameters, but apparently not as sensitive to the nature of the applied threshold. The value of $\psi(z)$ at a given redshift, however, does depend on the threshold value being considered, both for $S^*$ and metallicity.

\begin{figure*}
	\includegraphics[scale=0.47]{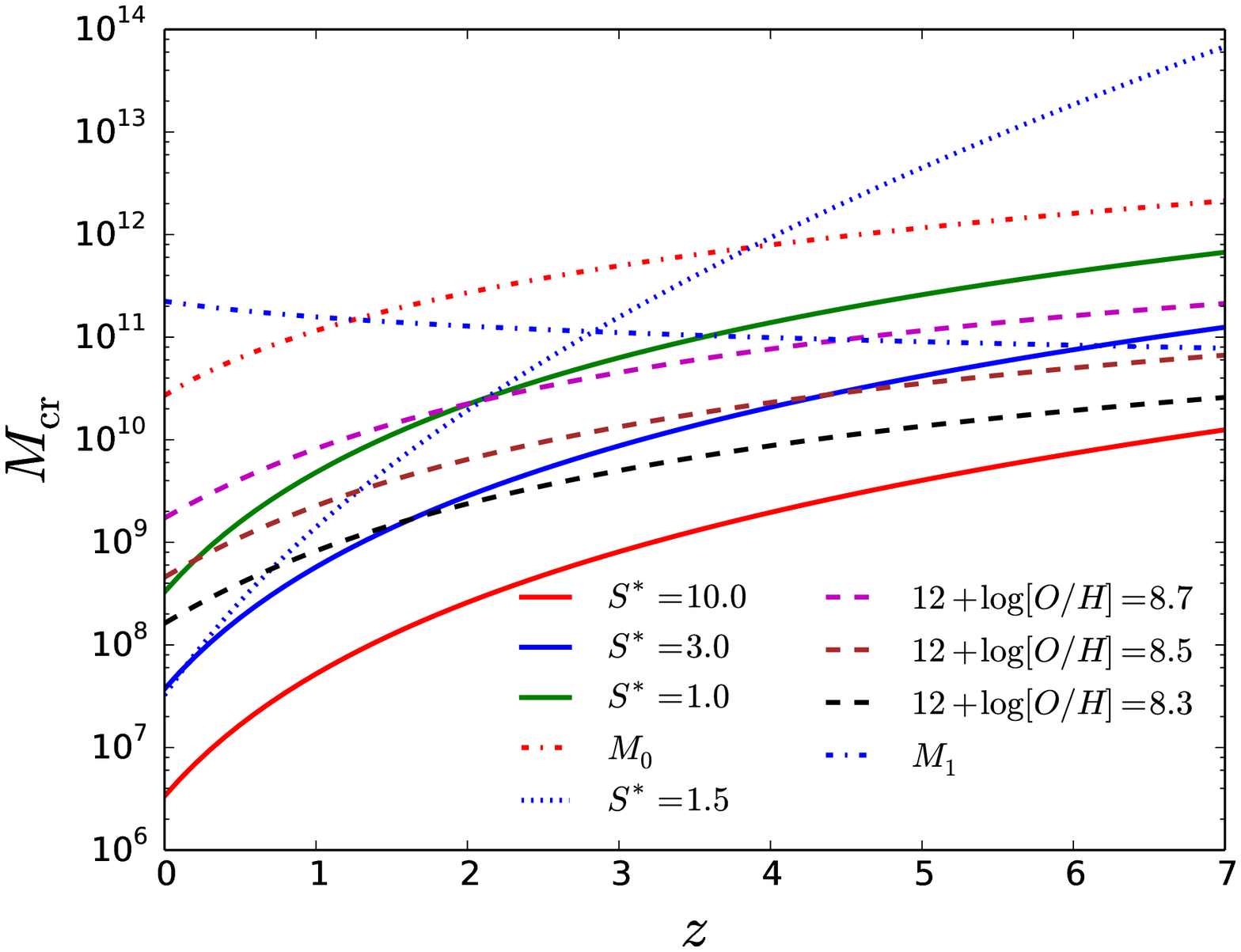}	
	\includegraphics[scale=0.47]{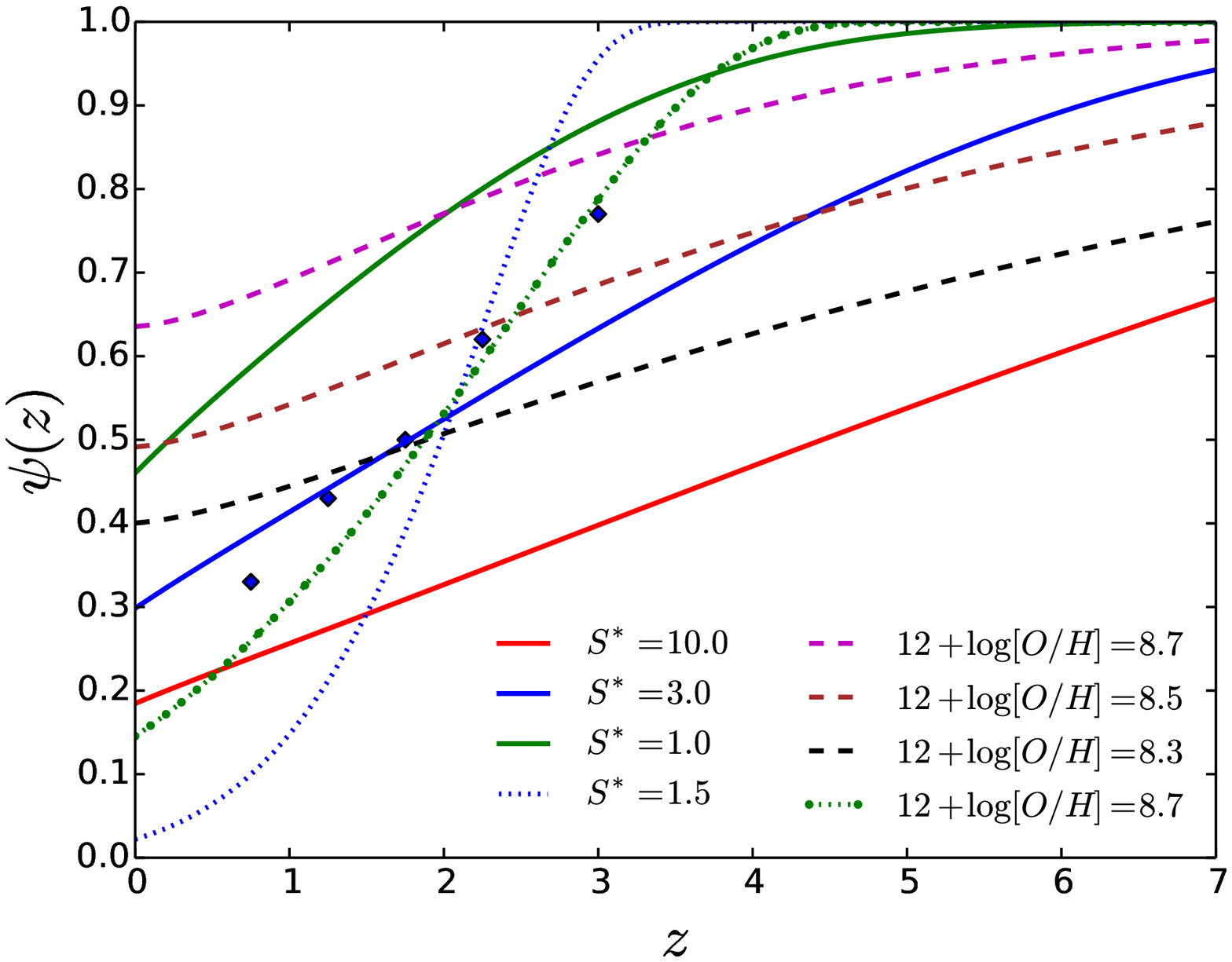}	
	\caption{(Left panel) Same as Fig.~\ref{fig:mcr}, with the addition of a critical mass obtained from the star formation rate from of \cite{Speagle2014} (eq.~\ref{eq:sfrm2}, dotted curve). The critical mass is calculated corresponding to a threshold value of $S^*=1.5$ Gyr$^{-1}$. See text for further discussion. (Right panel) Same as Fig.~\ref{fig:psi} with the galaxy mass function from \cite{Fontana2006} and star formation rate of \cite{Speagle2014}. A high metallicity threshold now gives a reasonable rembleance to the data, whereas the SFR threshold results in a too steep evolution.}
	\label{fig:psi2}
\end{figure*}

To further explore the effects of a different SFR-stellar mass description, we show in Fig.~\ref{fig:psi2} the results obtained with the rate of \cite[][eq.~20]{Speagle2014},
\begin{equation}
SFR(M,z)=S_0(M)(1+z)^{a(\log M)}
\label{eq:sfrm2}.
\end{equation} 
Here, $S_0(M)=10^{b(\log M)}$, and the stellar-mass dependent coefficients $a(\log M)$ and $b(\log M)$, come from the fit of \cite{Speagle2014}. 
\footnote{We choose their fit \#131 in Table 8, which includes high-redshift data extending to $z=5$, while the mass range is $10^{9.2-11.2}M_\sun$. We assume that this fit can be extrapolated to higher and lower masses and that it is valid throughout our redshift range.}  We adopt this version rather than their time-based parameterization, for ease of comparison with earlier works and our eq.~(\ref{eq:sfrm}).

In the left panel of Fig.~\ref{fig:psi2}, we plot the critical mass as a function of redshift, calculated from eq.~(\ref{eq:sfrm2}), for a specific rate threshold of $S^*=1.5$ Gyr$^{-1}$. This shows a steeper growth with increasing $z$ than previous critical masses. Recall that higher or lower specific rate thresholds will shift this curve down or up, respectively. Here, high mass galaxies will not be hosting a GRB out to a redshift of $z\approx 3$, in agreement with our earlier discussion.

The combination of the \cite{Fontana2006} mass function and the \cite{Speagle2014} SFR relation (eq.~\ref{eq:sfrm2}) gives the (blue) dotted and (green) dashed-dotted curves in Fig.~\ref{fig:psi2} (right panel). For the metallicity threshold (dashed-dotted), the star formation fraction, $\psi(z)$, is very similar to the corresponding curve in Fig.~\ref{fig:psi} (right panel), but now a higher threshold value of $12+\log[O/H]=8.7$ is required. As the upper integration limit in eq.~(\ref{eq:psi}) does not depend on the galaxy relations, comparison of the metallicity threshold curves in the right panels of Figs.~\ref{fig:psi} and \ref{fig:psi2} implies that the redshift dependence of $\psi(z)$ is more sensitive to the galaxy mass function than the SFR relation. 

Adopting a specific SFR threshold of $S^*=1.5$ Gyr$^{-1}$ instead, we find that $\psi(z)$ evolves much more steeply with the redshift (dotted curve), a reflection of the steeper redshift dependence of the corresponding critical mass (upper integration limit; Fig.~\ref{fig:psi2}, left panel). In fact, it is difficult to account for the SHOALS data with an $S^*$ threshold model using the \cite{Speagle2014} fit to the SFR-stellar-mass relation. 

We have not attempted to fit or constrain our adopted galaxy stellar-mass and SFR models with the SHOALS data points, as the galaxy functions parameters we use are obtained from large surveys while the SHOALS fractions are based on too small a sample to meaningfully constrain the galaxy relations.
It is clear, however, that the resulting star formation fraction is dependent on the details in the galaxy mass function and the galaxy stellar-mass-SFR relation. With an increased sample size, host galaxies of GRBs may provide additional constraints on those relations in the future.


\section{Discussion and Conclusions}
\label{conclusions}

We have used models of the galaxy stellar-mass function, the galaxy SFR function, and galaxy mass-metallicity relation to estimate the fraction of galaxies likely to host a GRB, as well as their contribution to the cosmic SFR. We impose a threshold value on either metallicity or specific SFR rate that translates into an upper mass limit on the host galaxies. Adopting analytic functions for the galaxy mass relation and SFR relation (eqs.~\ref{eq:phi} and \ref{eq:sfrm}), the SHOALS data can be
accounted for by either $S^*$ or metallicity thresholds, although the latter requires a metallicity threshold rather lower than that inferred from the data \citep{Perley2016b}. This conclusion, however, depends on the mass and SFR functions. By replacing eq.~(\ref{eq:sfrm}) with the description from \cite{Speagle2014} combined with the mass function from \cite{Fontana2006}, we find that a a high metallicity threshold may also explain the trend in the SHOALS data. We have also explored other combinations, such as using eq.~(\ref{eq:sfrm2}) with the \cite{DA2008} mass relation (eq.~\ref{eq:phi}), but this does not provide a likely combination to explain the data.

With our analytic choice of the SFR-stellar mass relation and the galaxy stellar-mass function, we find that the results presented here depend most strongly on the low mass slopes, $\beta$ and $\gamma$, and the mass scales, $M_0$ and $M_1$. In fact, $\gamma$ is initially assumed here to be constant, whereas \cite{Fontana2006} and \cite{Davidzon2017}, both find it to be redshift dependent. The detailed $z$ behavior of the low mass slopes as well as $M_0$ and $M_1$, is influenced by the galaxy
sample and redshift range used to estimate these parameters. Our choice of parameterizations and the functional form adopted for the SFR and galaxy stellar-mass functions is to ease comparison with earlier work and lead to a simple analytical expression for $\psi(z)$. As shown in Figs.~\ref{fig:psi} and \ref{fig:psi2}, different choices will lead to different redshift behavior of $\psi(z)$ for both types of thresholds and are easy to implement within our approach. An updated metallicity
parameterization is, however, needed if the formulation of \cite{Savaglio2005}, is to be replaced, a non-trivial task given the uncertainties in high-redshift galaxy selection ($z>3.5$) and lack of optical spectral lines for metallicity determination \citep{Maiolino2019}. 

Comparing the star formation fractions obtained with two different SFR relations (eqs.~(\ref{eq:sfrm}) and (\ref{eq:sfrm2})), combined with the mass function from \cite{Fontana2006} and a metallicity threshold, we find that these have very similar redshift behavior and follow the host fractions rather well. The corresponding metallicity threshold value is, however, considerably higher using the parameterization from \cite{Speagle2014}. Contrasting this with the results of the steeper critical
mass evolution obtained from eq.~(\ref{eq:sfrm2}) with an $S^*$ threshold, it now shows a much steeper star formation fraction evolution than seen in the SHOALS data. A number of different functional forms of the galaxy relations and fits to them are available in the literature, but an extensive exploration of these and the resulting star formation fractions is beyond the scope of this work. It is clear, however, that they will all show similar redshift behavior of $\psi(z)$ as we have explored here, but will of course differ in the details.

Clearly, there is room for improvement, but large galaxy surveys to high redshifts are essential for that. For example,  \cite{Tomczak2014} consider the redshift range $0.2\lesssim z \lesssim 3$, and fit the stellar-mass function with both single and double Schechter functions \citep[see also][]{Davidzon2017}, while \cite{Tomczak2016} extend the study up to $z\lesssim 4$, but use a different functional form for the SFR-stellar-mass relation. \cite{Salmon2015} find no evidence for a redshift
evolution of the SFR-stellar-mass slope in the range $4\lesssim z\lesssim 6$ from CANDELS. \cite{Bhatawdekar2018} find no evolution in $M_1$ over the range $z \sim 6-9$, but a steepening of the slope for $z \gtrsim 6$. 

\cite{Modjaz2019} assume and argue for low metallicity being the more fundamental parameter for hosts, albeit for a low-redshift sample ($z<0.2$), while \cite{Perley2016b} conclude that an evolving metallicity threshold is not seen in their data. In fact, their data is consistent with a non-evolving metallicity threshold that is close to solar, considerably higher than single star progenitor models predict \citep[e.g.][]{MacFadyenWoosley1999, LangerNorman2006}. We find, for our analytical choice
of galaxy relations,  that a low metallicity threshold can reasonably account for the SHOALS star formation fractions although it is, at the same time, at tension with the SHOALS inferred high metallicity threshold. A different mechanism or threshold may therefore be more relevant for a GRB production in a given galaxy, and we propose here that this is provided by a high specific SFR. 

Our assumption of a redshift-independent specific rate threshold may also prove to be too strict. Assuming instead that the threshold evolves as a power law in $(1+z)$ could actually result in a weaker redshift dependence of the corresponding $M_{\rm cr}$, as easily seen from equation~(\ref{eq:ssfr}), but depending on the sign of the power-law exponent. This could in turn lower the fraction of galaxies at high redshift contributing to the GRB rate, helping to reduce the difference in SFR density
as inferred from hosts, as compared to galaxy surveys.  

Samples even larger than SHOALS will likely be required to fully decide on the relative importance of $S^*$ vs.~metallicity in GRB production. We feel, however, that a better comprehension and parameterizations of the redshift evolution of the galaxy stellar-mass function and the SFR-stellar-mass function is even more crucial for a full understanding of the problem. In fact, host galaxies of GRBs may then provide additional constraint on these functions.

\section*{Acknowledgments}

I thank the anonymous referee for very helpful comments that broadened our results. This work was supported in part by the University of Iceland Research Fund. I thank P\'all Jakobsson and K\'ari Helgason for useful discussions and Kasper Elm Heintz for critical comments on the manuscript. I thank NORDITA, Stockholm, for hospitality during the initial stages of this work and I gratefully acknowledge the National Institute of Polar Research (NIPR) in Japan for hospitality during the final writeup. I thank Gu{\dh}j\'on H.~Hilmarsson and M\'oei{\dh}ur {\TH}orvalds\'ottir for assistance with the initial numerical calculations. Finally, I thank Dan Perley and coworkers for a very thorough and illuminating analysis of the SHOALS sample. 



\end{document}